# A Model for the Adoption Process of Information System Security Innovations in Organisations: A Theoretical Perspective

**Mumtaz Abdul Hameed**
Technovation Consulting and Training Private Limited
33, Chandhani Magu
Male'. Maldives
Email: mumtazabdulhameed@gmail.com

**Nalin Asanka Gamagedara Arachchilage**
Australian Centre for Cyber Security
University of New South Wales (UNSW Canberra)
The Australian Defence Force Academy
Australia
Email: nalin.asanka@adfa.edu.au

## Abstract

In this paper, we develop a theoretical model for the adoption process of Information System Security innovations in organisations. The model stemmed from the Diffusion of Innovation theory (DOI), the Technology Acceptance Model (TAM), the Theory of Planned Behaviour (TPB) and the Technology-Organisation-Environment (TOE) framework. The model portrays Information System Security adoption process progressing in a sequence of stages. The study considers the adoption process from the initiation stage until the acquisition of innovation as an organisational level judgement while the process of innovation assimilation and integration is assessed in terms of the user behaviour within the organisation. The model also introduces several factors that influence the Information System Security innovation adoption. By merging the organisational adoption and user acceptance of innovation in a single depiction, this research contributes to IS security literature a more comprehensive model for IS security adoption in organisation, compare to any of the past representations.

# 1　Introduction

Information is a key asset within an organisation and it needs to be protected (Arachchilage et al. 2013). It is profoundly important for an organisation to preserve information and computer resources (hardware, software, networks, etc.) collectively be referred as Information System (IS) assets against malicious attacks such as unauthorized access and improper use. In effect, safeguard of IS assets is a widespread concern in many organisations. A significant work has been done to develop and implement systems that would protect IS assets against malicious attacks (Yeh and Chang 2012). There are numerous measures available that provides protection for IS assets, including antivirus, firewall, filters, Intrusion Detection System (IDS), encryption tools, authorization mechanisms, authentication systems and proxy devices. In addition, computer security education needs to be considered as a means to combat against ISs threats (Arachchilage and Love, 2013; Arachchilage and Love, 2014; Arachchilage et al., 2016).

Research on the preservation of IS assets falls under the theme of IS Security. Consequently, the main focus of IS security is the implementation of strategies to protect and safeguard IS assets from vulnerabilities (Alshboul 2010). However, ensuring successful adoption and implementation of IS security in an organisation is a complex practice. It requires the full commitment of the staff and management. Previous scholarly contributions have constantly argued that the weakest link in any IS security plan or procedure is the computer users themselves (Wynn et al. 2012; Arachchilage et al. 2016). Security breach incidents have cost organisations, millions of dollars in lost and in the majority of these cases, fingers are pointing towards employee negligence and non-compliance (Herath and Rao 2009).

Adoption of IS security measures by the individuals and organisations is exceptionally low, considering the efforts put in for developing and implementing such systems (Lee and Kozar 2005; Tuncalp 2014). Hence, it is important to understand what cause users accept or reject the organisations' IS security measures (Jones et al. 2010). Examining the processes involved in the adoption of IS security strategies is fundamental for ensuring successful adoption process in organisations. As far as one can tell from the literature, there is no such a model that fully explains the IS security adoption process in organisations. However, research on IS innovation has introduced models, theories and frameworks related to the adoption and diffusion of IS innovations in organisations (Hameed et al. 2012a). Fundamentally, IS scholar's defined innovation as an idea, a method, a product, a program or a technology that is new to the adopting unit (Cooper and Zmud 1990; Damanpour 1991; Hameed et al. 2012a). On this ground, IS security may rightfully be considered as an IS innovation. Hence, theories based on innovation adoption may explicitly be applied in an empirical study on IS security adoption process.

Lack of proper IS security adoption model is the main hindrance for organisations from attaining a successful adoption process. This research attempts to examine IS security adoption process in organisations which includes organisational adoption process and the user acceptance of innovation. Therefore, we aimed to theoretically construct a model for IS security innovation adoption process in organisations. To this end, we explore past literature on the stages of innovation adoption, theories of innovation adoption, models of technology acceptance and popular frameworks developed by researchers for an organisational adoption with factors considered to influence IS innovation adoption. This study, then utilised appropriate concepts and relationships of prominent IS innovation adoption theories and user acceptance models to explain the organisational adoption of IS security practices. In addition, the study also identified a number of factors that influence innovation adoption in different contexts germane to the IS security adoption process. Based on the literature search, the current study combined the most suitable models and theories of innovation adoption with frameworks used in assessing organisational adoption perceptions. The integrated illustration of these models could very well be used to examine the adoption process and user acceptance of IS security innovations in organisations.

The study focuses on IS security adoption in organisations. The research makes three main contributions to the theory and practice of IS security. First, it draws upon and synthesize the rich literature in IS innovation adoption theories and applied it in the context of IS security innovations. We develop a model based on a mixture of four theoretical perspectives of innovation adoption that methodically emphasizes IS security adoption in organisations. Secondly, the proposed IS security adoption model encompasses both the organisational adoption process and user acceptance of innovation. Most past studies on IS security adoption only examine the processes of adoption of IS security innovation until the acquisition of innovation with no assessment on whether the innovation grows to be part of their regular practice (Lee and Kozar 2005; Safa et al. 2015). On the other hand, studies on user acceptance have only examined the behaviour and attitude of individuals accepting an IS security innovation (Li 2015; Salleh et al. 2015). Previous studies have rarely examined the adoption process and user acceptance of IS security innovations in organisations collectively. Combining the organisational adoption process and user acceptance of innovation in a single model allows depicting the overall adoption process more comprehensively compare to any of the past IS security adoption frameworks put forward. Thirdly, the proposed model has introduced several determinants that may influence IS security innovation adoption in organisations. The suggested association between various technological, organisational, environmental, and user acceptance characteristics for IS security adoption provides a rich ground for prospective research. Furthermore, the IS security adoption model proposed in this study provides important implications for practice as well as further research.

The remainder of this paper is organised as follows. The 'Theoretical Background' section illustrates the basics of IS security and different IS security innovations in organisations. The 'Methodology' section, briefly discusses the methods used to identify theoretical elements for the development of a model for the process of IS Security innovation adoption in organisations. In the subsequent section 'Models of IS Innovation Adoption', we identify the most prominent models of innovation adoption research and the fundamentals of innovation adoption in organisations. The fifth section presents the models use in the past IS security innovation adoption studies. In the 'IS Security Innovation Adoption Model' section, we discuss the relevance of innovation adoption models to explicate IS security innovation adoption process. Also, in this section we identify a number of factors investigated in the past literature that hypothetically influence adoption of IS security innovation. Furthermore, in this section, we present the proposed model for IS security innovation adoption in organisations. Finally, conclusion and future research regarding the model were presented in section 7.

## 2   Theoretical Background

Organisations are becoming more and more open to both internal and external threats to their ISs (Jones et al. 2010). Thus, IS security is still a prevalent issue among experts as well as users (Safa et al. 2015) and the risk of computer crimes has become a growing concern for many companies. The main challenge for organisations IS security is to protect unauthorized access of information sources and to defend computer resources against malicious attacks such as phishing, virus, malware, spyware, botnets or Distributed Denial Of Service (DDOS) attacks (Feruza and Kim 2007).

Increasing reliance on ISs in organisational operations has obliged the management to invest more on improving their IS reliability. ISs need to be secure if they are to be reliable. The most important IS security concern for organisations is to protect the Confidentiality, Integrity and Availability (CIA) of their information (Feruza and Kim 2007). Safeguard and the management of these three attributes of information, essentially accounts the entire issue of IS security in organisations (Cooper 2009). Confidentiality means that the information should be kept secret and only the people who are authorized to access may use it (Feruza and Kim 2007). Integrity of information refers to the correctness and completeness of information as well as prevention of improper and unauthorized modification of information, moreover, availability is related to the ease with which information is accessible to authorized users whenever required (Jones et al. 2010). As a safeguard measure, organisations are required to implement policies, practices and technologies that protect against unauthorized access, use, disclosure, disruption, modification or destruction of information (Feruza and Kim 2007). Although there is no such a standard mechanism to completely safeguard all of the IS assets of an organisation, a handful of measures can be put in practice to limit the number of attacks (Feruza and Kim 2007; Albuquerque Junior and Santos 2015).

A comprehensive range of security measures in the form of physical controls, procedural controls and technical controls would thwart almost all forms of security breaches to ensure Confidentiality, Integrity and Availability of information in an organisation (Feruza and Kim 2007). Physical controls are implemented to monitor and control the environment of the work place and computing facilities, whereas, procedural controls are aimed to change the people's behaviour, put across in the form of written policies, procedures, standards and guidelines to adhere within the organisation (Albuquerque Junior and Santos 2015). Technical controls utilise software and hardware to monitor and control access to information and computing facilities (Feruza and Kim 2007).

Any physical, procedural or technical security control put in place in an organisation to protect information and computer resources may possibly be characterized as IS security innovation. Damanpour and Wischnevsky (2006) defined innovation as the possession of ideas, systems, practice, products or technologies that are new to the adopting organisation. What's more, adoption of innovation is a process that results in the introduction and use of products, processes, or practices that are new to the adopting organisation (Damanpour and Wischnevsky 2006; Hameed et al. 2012a). Damanpour (1991) defines adoption of innovation as the generation, development and implementation of new initiatives or activities. Hence, implementation and the use of physical, procedural or technical security control may be considered as the adoption of IS security innovation in an organisation.

Correct IS security measures in organisations have long been recognized, however, the empirical research in this area is still at its early stage. Although there are a number of IS security innovations available, an organisation can only benefit if those innovations are adopted and implemented properly. The main hindrance for organisations from attaining a successful implementation of IS security innovation is the lack of appropriate models of IS security adoption. Therefore, this research attempts to examine IS security adoption process in organisations.

## 3   Research Methodology

The study initially performed a literature search to identify theoretical models utilised in examining adoption and user acceptance of IS innovations. Based on this search result, the study then identified the most commonly used innovation adoption and user acceptance models. The IS security adoption studies that used IS innovation adoption models in their empirical investigations were then selected. The IS security literature extracted includes studies conducted for both individual and organisational contexts. The most prominent innovation adoption models used in IS security adoption were then drawn together, to synthesize the conceptual model presented in this study. In addition, we extracted the factors from different categories that were examined in the IS security adoption literature.

## 4   Models of IS Innovation Adoption

A significant amount of research has been conducted in examining the process and the factors influencing the adoption and user acceptance of innovations in organisations. However, there is no organisational innovation adoption theory in existence for researchers to utilise (Hameed et al. 2012a). Hitherto, researchers have been utilising several theories and theoretical models to explain the adopter's attitude, innovation adoption behaviour and various determinants in different contexts of IS adoption. In addition, innovation adoption research has introduced several theoretical models related to the adoption and user acceptance of innovation in organisations (Hameed et al. 2012a).

The most common theoretical models used to examine adoption and user acceptance of innovation are Diffusion of Innovation Theory [DOI] (Rogers 1983), Perceived Characteristics of Innovation [PCI] (Moore and Benbasat 1991), Theory of Reasoned Action [TRA] (Fishbein and Ajzen 1975), Theory of Planned Behaviour [TPB] (Ajzen 1991), Technology Acceptance Model [TAM] (Davis 1989), Technology Acceptance Model 2 [TAM2] (Venkatesh and Davis 2000), Technology Acceptance Model 3 [TAM3] (Venkatesh and Bala 2008), Technology, Organisation, Environment [TOE] Model (Tornatsky and Fleischer 1990) and the Unified Theory of Acceptance and Use of Technology [UTAUT] model (Venkatesh et al. 2003). Amongst all of these models, DOI, TAM, TRA, TPB and TOE have been widely used in innovation adoption research (Hameed et al. 2012a). DOI, TAM, TRA and TPB are primarily utilised in examining the user behaviour of innovation adoption and TOE framework has widely been exploited in organisational level studies of IT innovation adoption.

Innovation adoption processes in an organisation are considered to be successful only if the innovation is implemented in the organisation and individuals continue to use the innovation over a period of time (Gopalakrishnan and Damanpour 1997; Hameed et al. 2012a). Based on this perception, the model presented by Hameed et al. (2012a) for IT innovation adoption for organisations considered both organisational level analysis and individual level assessment. In addition, the process of adoption of innovation in organisations has been categorized as a stage-based process (Rogers 1995). Researchers have described the process of adoption of innovation into a number of sequences of stages. According to Hameed et al. (2012a), the cycle of stages illustrated by different research falls more or less into the initiation, adoption-decision and implementation stage. These three phases of initiation, adoption-decision and implementation are more often referred to as pre-adoption, adoption-decision and post-adoption in the IS literature.

## 5   Research on IS Security Innovation Adoption

The aim of this study is to develop a conceptual model for IS security adoption that includes the process of adoption and user acceptance of IS security innovations in organisations. A search in literature confirmed that there is hardly any distinct theoretical model with the aim of explaining IS security adoption. IS security research mainly utilised IT innovation adoption and user acceptance models (Lee and Kozar 2005; Claar and Johnson 2012). In addition, researchers have applied models from other disciplines such as health belief model to examine user behaviour of IS security adoption (Ng et al. 2009; Claar and Johnson 2010). These studies propose a number of models and a wide range of determinants that influence an individual's decision to adopt IS security innovations.

For example, Lee and Kozar (2005) used TPB model to identify the factors influencing the user adoption of anti-spyware systems. The research examines the influence of three constructs of TPB i.e. attitude, social influence and Perceived Behavioural Control (PBC) for anti-spyware adoption of individuals. Similarly, in a review to observe the user behaviour to conscious care behaviour in the domain of information security, Safa et al. (2015) utilised TPB model. Lee and Kozar (2008) applied DOI and TPB model for an empirical investigation of anti-spyware adoption of computer users. The study investigates the attributes of DOI and TPB for user's anti-spyware adoption intention. For a research to examine the factors that influence employee acceptance of IS security measures, Jones et al. (2010) extended the TAM. Some past studies have also suggested certain organisational factors that influence adoption of IS security innovation. In a study to measure and identify factors influencing online security performances, Li (2015) use the TOE framework. Likewise, to explore the security determinants in big data solutions adoptions, Salleh et al. (2015) adopted the TOE structure as the main conceptual research framework.

By and large, IS security adoption studies have examined individual attitudes and behaviour towards innovation (Lee and Kozar 2008; Jones et al. 2010; Safa et al. 2015). Research on IS security rarely considers the adoption process at the organisational level. Meanwhile, innovation adoption literature suggests that researchers have been utilising several theories and theoretical models that explain the adopter's attitude and organisational innovation adoption's behaviour to examine different types of innovation such as IS security. As a result, a suitable model or models in the IS domain that is general enough may be exploited and perhaps be sufficient to explain IS security adoption in organisations. Indeed, a number of studies have introduced adoption and user acceptance models in the organisational context for various other innovations (Hameed et al. 2012b; Hameed and Counsell 2014a).

Most of the research on innovation adoption of organisational surrounding conducts their analysis by integrating innovation adoption and user acceptance theories with frameworks that consists of determinants that are relevant to the study context. For example, Teo et al. (2009) empirically examined adopters and non-adopters of e-procurement in Singaporean organisations, incorporating two innovation adoption theories and a framework consisting determinant of TOE model. What's more, Hameed et al. (2012a) proposed a more general IT innovation adoption model for organisations by combining innovation adoption and user acceptance theories, and major frameworks used in IT innovation studies. Their model is a combination of DOI, TRA, TAM, TPB and a framework that consists of determinants of TOE and Chief Executive Officer (CEO) characteristics. The model exploited DOI model and TOE framework with CEO characteristics to illustrate the organisational adoption process until the acquisition of innovation and TRA, TAM and TPB were utilised to construct user acceptance of innovation. Here, TOE framework takes account of the various determinants relevant to IS innovation adoption in organisations.

Hence, a theoretical model for the adoption of IS security innovation in organisations may consist of a combination of innovation adoption and user acceptance theories jointly with contextual frameworks of IT innovation adoption. It is evident from the literature that previous scholarly IS security adoption contributions have on no account addressed organisational adoption process and user acceptance of innovation in a single investigation.

# 6  IS Security Innovation Adoption Model

Establishing the views of IS innovation literature and consistent with the model presented by Hameed et al. (2012a), the model we present in this study discusses IS security innovation adoption for organisations as a two level adoption proceeding, an organisational level investigation and individual or user level assessment. The moment organisation starts pursuing the knowledge of the IS security innovation until the actual acquisition of innovation is regarded as an organisational level adoption process. The user acceptance of IS security innovation along with the actual use of innovation is classified as individual level or user exhibit adoption process.

To go along with the foundations of much of the previous research in IS innovation adoption (Pierce and Delbecq 1977; Rogers 1995; Hameed et al. 2012a) this study considers IS security innovation adoption process in organisations as a three stage process of pre-adoption, adoption-decision and post-adoption. The study deems that the pre-adoption stage consisting of activities related to recognizing a need, acquiring knowledge or awareness, forming an attitude towards the innovation and proposing innovation for adoption (Rogers 1995; Gopalakrishnan and Damanpour 1997). The adoption-decision stage described by Meyer and Goes (1988) reflects the decision to accept the idea and evaluates the proposed ideas from a technical, financial and strategic perspective, together with the allocation of resources for its acquisition and implementation. The study also considers the post-adoption stage, which involves the acquisition of innovation, preparing the organisation for the use of the innovation, performing a trial for confirmation of innovation, acceptance of the innovation by users and continued actual use of the innovation (Rogers 1995).

We developed the IS Security Adoption Model by replicating the theories of IS innovation adoption, at the same time, being tightly consistent with prior research applying these theories for different IS innovation perspective. Based on innovation adoption literature, the study draws together a conceptual model for IS security innovation adoption by integrating multiple theoretical models of innovation adoption and user acceptance of IS with the popular frameworks. The model is a combination of DOI, TAM, TPB models together with the TOE framework.

DOI is the most generally accepted model for identifying the main characteristics of IS innovation adoption (Premkumar and Roberts 1999; Hameed et al. 2012a). Although the model has a solid theoretical foundation, the model only explicates the individual level adoption process and it does not include the post-adoption behaviour of the innovation adoption process. Hence, DOI alone cannot be used to fully explain IS security innovation adoption in organisations. TAM and TPB provide a basis for presenting the post-adoption behaviour of innovation adoption. Consequently, TAM and TPB, singly and jointly been used in empirical investigations to predict and explain user acceptance of IS innovation (Hameed and Counsell 2014b). TPB complements TAM's constructs at the same time TPB explanatory and predictive power enhances further by integrating with TAM (Awa et al. 2014). TAM only account for behaviours where user of innovation is mandated. On the other hand, TPB constructs would allow predicting for both volitional and non-volitional conditions (Hameed et al. 2012a). Hence, combining DOI with TAM and TPB helps us to derive a model that reflects pre-adoption, adoption-decision and post-adoption stages of IS security innovation adoption. DOI, TAM and TPB models have been successfully exploited and effectively been used in explaining and predicting either the adoption or user acceptance of IS innovations for individual and organisational context (Hameed et al. 2012a). However, DOI, TAM and TPB are individual level adoption models. Therefore, researchers have combined DOI, TAM and TPB with a contextual framework to address organisational level innovation adoption process (Hameed et al. 2012a). The TOE model has been extensively adapted to identify factors influencing the adoption of IT innovations in organisations. Thus, an integrative model consisting of DOI, TAM, TPB and TOE would fully explain IS security innovation adoption in organisations.

The proposed model uses the constructs of TAM and TPB to account for the user acceptance of IS security innovation. Hence, the user acceptance attributes of TAM and TPB affects the IS security adoption process at the post-adoption stage. Use of DOI and TOE in the proposed model could successfully explicate the adoption process at organisation perspective. In light of the technology, organisation and environment attributes that facilitate the adoption, both DOI and TOE competently elucidate pre-adoption and adoption-decision stages of IS security adoption.

Technological context of TOE model describes that adoption depends on the number of technologies inside and outside of the firm. The importance of technology attributes for the adoption and implementation of IS and perception of innovation influencing the pre-adoption and adoption-decisions have been documented in the IS literature (Rogers 1983). Specific characteristics of innovation are examined as factors that explain innovation adoption in organisations. DOI theory provides a set of innovation attributes that may affect the adoption decision (Rogers 1995). The organisational context of TOE model has been the most frequently examined attributes in adoption of IS innovations in organisations. Researchers have advocated the primary importance of organisational determinants compared to other contexts as predictors for innovation adoption (Damanpour 1991; Hameed et al. 2014a). The environmental context of TOE model relates to facilitating and inhibiting factors in areas of operations. Organisations are adopting innovation in response to an external demand or to achieve an advantage of an environmental opportunity (Hameed and Consell 2012). IS has not only been used for internal needs; instead, organisations often communicate with customers, suppliers and other trading partners. Hence, environmental factors are increasingly being studied in innovation adoption studies.

Constructs of TAM and TPB contribute most towards user acceptance attributes. The two attributes of TAM, perceived usefulness and perceived ease of use were key determinants of user IS acceptance (Hameed and Counsell, 2014b). The PBC factor of TPB was found to be significant and sub-constructs of PBC (computer self-efficacy and facilitating conditions) which determined non-volitional behaviour was also found to be significant characteristics.

For the IS security adoption model, we extracted technology, organisation, environment and user acceptance attributes that were examined in the past IS security literature. Lee and Kozar (2005) found that user acceptance attributes of attitude, social influence, PCB and image together with technology related factors, including relative advantage, compatibility, visibility, and trialability plus the computing capacity of the organisation significantly influence anti-spyware systems adoption. Lee and Kozar (2008) examined anti spyware software adoption and found that user acceptance attributes of attitude, subjective norms, self efficacy, image and PCB along with technology related factors, including relative advantage, compatibility, visibility and trialability as well as computing capacity of the organisations, as important determinants. Jones et al. (2010) examined TAM attributes and found perceived usefulness; perceived ease of use and subjective norms had a significant effect on intention to use IS security measures. To verify IS conscious care behaviour formation in organisations, Safa et al. (2015) found an important relationship of user attitude, subjective norms, self-efficacy, IS awareness, experience and policy, and organisation policy. Li (2015) verified that three TOE factors, namely: importance of IS security, firm size and the existence of government regulation, all predicts online security performance in organisations. Similarly, Salleh et al. (2015) utilised the TOE framework to explore security determinants of big data solutions. Technological factors listed in the study were perceived complexity and perceived compatibility, organisational factors considered were top management support, IS culture and organisational learning culture, finally environmental factors included were security regulatory concerns and risk of outsourcing.

Among the attributes examined in the IS security literature, we considered the factors that showed a significant relationship with IS security adoption. Hence, for this study, we considered the following user acceptance and TOE factors. Relative advantage, compatibility, complexity, visibility and trialability were included in terms of technology characteristics. For organisational characteristics, the model proposed top management support, organisational size, IS awareness, IS experience, organisational policy, computing capacity, IS culture and organisational learning culture. In the IS literature, computing capacity described by Lee and Kozar (2005) and Lee and Kozar (2008) is termed as IS readiness and we adopt this terminology in our model. In addition, IS awareness and IS experience described by Safa et al. (2015) is often described as a single attribute in the IS literature, namely: IS expertise (Hameed et al. 2012b). Hence, for the proposed model of IS security adoption, we refer both factors as IS expertise. Furthermore, according to Leidner and Kayworth (2006) IS culture is a variable that explains how social groups interacts with IS in the development, adoption, use and its management. Hence, both the factors IS policy and IS learning culture could simply be classified under IS culture. In our proposed model, we included IS culture as a determinant that accounts the perception of IS policy and IS learning culture. The recommended attributes of the environmental context include government regulation and risks of outsourcing. User acceptance determinants for the model were user attitude, subjective norms, perceived behavioural control, computer self-efficacy, perceived usefulness, perceived ease of use and image.

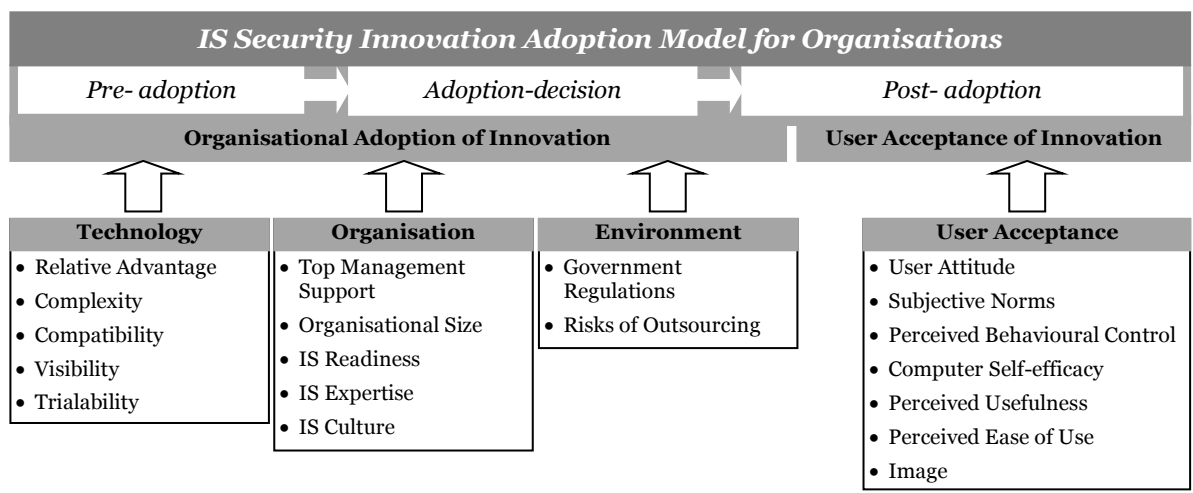

*Figure 1: The proposed conceptual model for IS security innovation adoption in organisations.*

Fig. 1 illustrates our proposed conceptual model for the IS security innovation adoption in organisations. Table 1 provides a description for each of the factors considered in the IS security innovation adoption model.

| Name | Description |
|---|---|
| **Technology Factor** | |
| Relative Advantage | The degree to which an innovation is perceived as being better than the idea it supersedes [Hameed et al., 2012a] |
| Complexity | The degree to which an innovation is perceived as relatively difficult to understand and use [Hameed et al., 2012a] |
| Compatibility | The degree to which an innovation is perceived as being consistent with the existing values, past experiences and needs of the users [Hameed et al., 2012a] |
| Visibility | The degree to which an individual observes others' adoption of the innovation [Lee and Kozar, 2005] |
| Trialability | The degree to which an innovation may be experimented within a limited basis [Hameed et al., 2012a] |
| **Organisation Factors** | |
| Top Management Support | The extent of the commitment of resources and support from the top management to the innovation [Hameed et al., 2012b] |
| Organisational Size | Number of employees within the organisation or total sales revenue [Hameed et al., 2012b] |
| IS Readiness | The degree to which innovation has the capacity to fit on to one's computer and networks [Lee and Kozar, 2005] |
| IS Expertise | Prior experience of innovation in term knowledge of individuals and within the organisation [Hameed et al., 2012b] |
| IS Culture | The behaviour in an organisation that contributes to the protection of information of all kinds [Salleh et al., 2015] |
| **Environment Factors** | |
| Government Regulation | Refer to government policies to promote IS security adoption and organisational concerns in ensuring compliance to security and data privacy regulations [Hameed and Counsell 2012; Salleh et al., 2015] |
| Risks of Outsourcing | Refer to the associated security and privacy risks that may result from an organisational decision to outsource their innovation adoption initiative [Salleh et al., 2015] |
| **User Acceptance Factors** | |
| User Attitude | The degree to which a person has a favourable or an unfavourable feeling about a behaviour [Lee and Kozar, 2008]. |
| Subjective Norms | The degree to which an individual perceives social pressure to adopt or not to adopt innovation [Lee and Kozar, 2005] |
| Perceived Behavioural Control | PBC is defined as "the perceived ease or difficulty of performing the behaviour [Lee and Kozar, 2008] |
| Computer Self-efficacy | The judgment of one's ability to use a computer and facilitating conditions [Hameed et al., 2012b] |
| Perceived Usefulness | Refers to the tendency to use or not to use an innovation to the extent it is believed that it will help or enhance an individual's ability to perform his or her job better [Jones et al., 2010] |
| Perceived Ease of Use | The degree a person believes an innovation is free of effort, he or she would be more likely to use and accept the innovation [Jones et al., 2010] |
| Image | The degree to which the adoption of an innovation enhances one's image as a technical and moral leader among his/her referents [Lee and Kozar, 2005] |

*Table 1. A list of factors included in the conceptual model for IS security innovation adoption model for organisations·*

# 7　Conclusion and Future Research

In this study, we developed and proposed a model for the process of IS security innovation adoption in organisations. The study integrated theoretical perspectives of IT adoption and user acceptance models and popular frameworks to build the integrative structure. The proposed model assessed the IS security adoption process, navigating from pre-adoption through adoption-decision and then post-adoption stages. The model described two levels of analysis, from the initiation stage until the acquisition of innovation was assessed as organisational process and the process of user acceptance of the innovation is analysed in terms of the behaviour of the individuals within the organisation. The structure is a combination of DOI, TAM, TPB and TOE framework. The study considered the IS security innovation adoption process to be successful only if the innovation is accepted and integrated into the organisation and the individual users continue using the innovation. The model exploited DOI and TOE framework to characterize the organisational adoption process until the acquisition of innovation and TAM and TPB to construct user acceptance of IT.

The study focused on IS security adoption in organisations. The contribution of the study includes an enhancement of our understanding of IS security adoption and implementation process in organisations. It draws upon and blends from the rich literature in IT innovation adoption theories and applies it in the context of IS security where, it has rarely been empirically investigated. To surmount the shortcomings of individual IS innovation adoption models such as the DOI and TAM; the proposed model combined a number of innovation adoption models. Merging different innovation adoption models allows the individual model to complement each other, hence, making structure of the proposed model more robust. Another important contribution of this research is that the proposed IS security model considers both the organisational adoption process and the user acceptance of innovation in a single illustration. Incorporating the organisational adoption process and the user acceptance of innovation in a single representation allows to depict IS security innovation adoption process more systematically.

In addition, the proposed model introduces several determinants that may influence IS security adoption in organisations, in particular, the association between various technological, organisational, environmental and user acceptance characteristics with IS security adoption. As this is a theoretical model developed using past studies, unavailability of any empirical insights of the model limits us to drawing causal implications of the findings.

However, the proposed model presented have considerable significance in understanding the process involved in the adoption of IS security innovation in organisations. Also, it allows to highlight the key steps to navigate to achieve a successful adoption of IS security innovations. Equally, the study provides researchers and practitioners with a set of factors that affect the adoption of IS security in organisations. It serves as a guideline for practitioners to identify and address the facilitating and inhibiting issues in the context of technology, organisation, environment and user acceptance attributes in the process of IS security adoption. Managers need to consider these issues when embarking on IS security adoption in their organisations.

The IS security adoption model proposed in this study provides important implications for practice as well as for further research. This study has a number of implications for managers and IS researchers. Managers can draw up this model and assess the condition of the IS security adoption process and possible factors that would lead to a successful adoption of IS security innovations in their organisations. In addition, managers can utilise the model to plan and prepare for the adoption process and establish smooth conditions for the user acceptance in the IS security implementation process.

The study has some limitations that need to be considered when interpreting the results. First, the methodology used and the methodological screening imposed for the inclusion and exclusion of studies may limit interpretation of the results. This is one of the limitations of our research that was unavoidable as the number studies that examined IS security adoption is very scarce. Second, limitation may be publication bias. As for any literature study, the review of studies may have been subjected to publication bias. However, with every effort to cover all the literature on IS security innovation adoption; the study may not be completely immune to publication bias. Another limitation is that this research obtained TOE and user acceptance attributes from a small number of studies. This could result in a narrow scope that does not adequately capture all TOE and user acceptance characteristics relating to IS security adoption.

In terms of future research, the proposed model identified the different factors that influence IS security adoption in the context of technology, organisation, environment and user acceptance behavioural characteristics. It gives no indication of the significance of each factor for different stages of IS security adoption. Researchers could extend this study by analysing the interaction between different characteristics and variables, since the impact of the attributes would be different at different stages of the IS security adoption process.

Furthermore, IT practitioners may utilise this model to investigate the factors influencing the adoption of IS security innovations in various demographic settings; the model could be tested by organisations from different sectors and different countries. Future research will refine the relationships using an empirical investigation to enable researchers to establish causal relationships.